\documentclass[11pt]{article}
\usepackage{moriondsilvia,epsfig}
\usepackage{graphicx}

\def\Journal#1#2#3#4{{#1} {\bf #2}, #3 (#4)}

\def\be{\begin{equation}}
\def\ee{\end{equation}}
\def\bea{\begin{eqnarray}}
\def\eea{\end{eqnarray}}

\begin{document}
\vspace*{4cm}
\title{FOREGROUNDS IN THE BOOMERANG-LDB DATA: \\
A PRELIMINARY rms ANALYSIS}

\author{ S. MASI$^1$, P.A.R. ADE$^2$, J. BOCK$^3$, A. BOSCALERI$^4$, 
B.P. CRILL$^5$, P. DE BERNARDIS$^1$, \\
K. GANGA$^5$, M. GIACOMETTI$^1$, E. HIVON$^5$, V.V. HRISTOV$^5$, 
A.E. LANGE$^5$,\\
L. MARTINIS$^6$, P.D. MAUSKOPF$^7$, T. MONTROY$^8$, 
C.B. NETTERFIELD$^9$, E. PASCALE$^4$, \\
F. PIACENTINI$^1$, S. PRUNET$^9$, G. ROMEO$^{10}$, J.E. RUHL$^8$, F. SCARAMUZZI$^6$ }

\address{
 1) Dipartimento di Fisica, Universita' La Sapienza, Roma, Italy\\
 2) Queen Mary and Westfield College, London, UK\\
 3) Jet Propulsion Laboratory, Pasadena, California, USA\\
 4) IROE-CNR, Firenze, Italy\\
 5) California Institute of Technology, Pasadena, CA, USA\\
 6) ENEA, Frascati, Italy\\
 7) Department of Physics, University of Cardiff, UK\\ 
 8) Department of Physics, University of Santa Barbara, CA, USA\\ 
 9)  Department of Physics and Astronomy, University of Toronto, Canada\\
 10) Istituto Nazionale di Geofisica, Roma, Italy 
}

\maketitle\abstracts{ We present a preliminary analysis of the BOOMERanG 
LDB maps, focused on foregrounds. BOOMERanG detects dust emission
at moderately low galactic latitudes ($b > -20^o$) in bands centered at
90, 150, 240, 410 GHz. At higher Galactic latitudes, we use the BOOMERanG 
data to set conservative upper limits on  the level of contamination at 90 
and 150 GHz. We find that the mean square signal correlated with the IRAS/DIRBE
dust template is less than 3$\%$ of the mean square signal due to 
CMB anisotropy.}

\section{ Introduction }

Millimeter-wave emission of interstellar dust can be 
an important contaminant of sensitive CMB anisotropy surveys 
(see e.g. \cite{ang}). However, experimental information 
on this subject is still coming from serendipitous
detections of dust in sensitive anisotropy experiments \cite{max1}
\cite{mas1}  \cite{mas2}  \cite{Leitch} \cite{Muk} , 
from the low-resolution surveys of 
COBE-DMR \cite{cobe1} and COBE-FIRAS \cite{cobe3},
and from extrapolation of the full sky surveys of 
IRAS and DIRBE \cite{Sch} \cite{Fink}.

BOOMERanG \cite{Lan} \cite{mas3} \cite{pdb1} is a sensitive 
millimeter wave telescope operating on a 
long duration balloon platform. The instrument is designed to take 
full advantage of long duration stratospheric flights, and
features a 1.3 m off/axis, low background, low sidelobes telescope,
and ultra sensitive bolometers covering 
4 frequency bands (90, 150, 240, 410 GHz) with resolution
ranging from 10 to 18 arcmin FWHM. 

In the 1998/99 long duration flight, BOOMERanG observed about 
2000 square degrees of the sky. About 1300 square degrees are 
at high galactic latitudes ($b < -20^o$), in a region of
sky with the lowest amount of dust emission
(constellations of Caelum, Doradus, Pictor, 
Columba, Puppis). The fluctuation of the 100 $\mu$m brightness mapped by IRAS 
is well below 1 MJy/sr in over 500 square degrees, 
as estimated from the DIRBE-recalibrated IRAS maps \cite{Sch}. 

In this paper we show that, in this region mapped 
by BOOMERanG, dust contamination in the CMB bands (90 and 150 GHz) 
is negligible with respect to the CMB signal. The 240 GHz band is 
partially contaminated by dust emission only in the lowest Galactic 
latitudes. The 410 GHz map is well correlated
with the DIRBE-recalibrated IRAS maps.

\section{ RMS Data Analysis }

The BOOMERanG maps used here have been obtained from the raw data using an 
iterative method\cite{pru} which greatly reduces the large scale 
artifacts due to low-frequency noise in the detector system, and correctly 
estimates the noise in the datastream. Low spatial frequencies (structures 
larger than $\sim 10^o$) have been removed from the maps. Healpix 7' 
pixelization has been used. 

\begin{figure}[!h]
\begin{center}
\includegraphics[height=7.5cm]{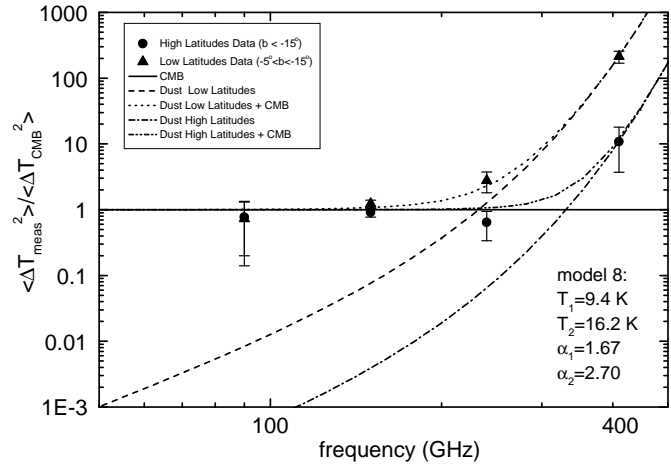}
\caption{Mean square temperature fluctuations measured by BOOMERanG at high
($b<-20^o$, circles) and low ($b>-20^o$, triangles) Galactic latitudes, 
normalized to the mean square CMB temperature fluctuations. 
The data are compared to a two components dust emission model
(Finkbeiner model number 8, thick lines). The extrapolations of the measured
dust fluctuations at 90 and 150 GHz are negligible with respect to the
measured CMB fluctuations.
\label{fig:rms}}
\end{center}
\end{figure}

In order to estimate the level of contamination in the 90, 150, 240 GHz channels, 
we proceed as follows:

1) We estimate the mean square value of sky signal fluctuations at low and 
high galactic latitudes. Uncorrelated detector noise is removed by 
correlating signals from different detectors ($A$ and $B$) in the same 
frequency band. This procedure is similar to the initial COBE-DMR analysis:
$$
<Sky^2>= {1 \over 4} [<(A+cB)^2>- <(A-cB)^2>]
$$
where $c$ is the slope of the best fit  $A = a + cB$.

2) We plot these fluctuations versus frequency, at high and low Galactic 
latitudes (see fig.1). Assuming a combination of cold and warm dust 
is appropriate for high Galactic latitudes \cite{Fink}. We normalize 
the best fit model in\cite{Fink} (model 8) to our 410 GHz mean square signal.
In fact, a visual comparison of the two maps shows that the two 
are very well correlated, and that our 410 GHz channel 
is dominated by dust emission in cirrus clouds.
From the figure is evident that
the model 8 dust spectrum fits quite well the 240 GHz point, and 
that dust is irrelevant at 90 and 150 GHz, 
producing less that 1$\%$ of the mean square fluctuation at 150 GHz. 
Even using a single-temperature dust, normalized at 410 GHz 
and with emissivity index $\alpha = 2$ we get very similar conclusions.

\section{ Correlations with IRAS }

The previous results are based on a spectral model for the extrapolation 
of dust emission at long wavelengths. In the following we want to 
find model-independent constraints on the level of dust fluctuations 
present in our bands. 

The best dust monitor we have is the IRAS/DIRBE map, so we pixel-pixel 
correlate our maps to a properly re-pixelized IRAS/DIRBE map, in order 
to get an estimate of the dust signal in our maps. Using the IRAS map 
as a dust template (rather than our own 410 GHz channel), we also remove 
any possible spurious instrumental correlation.

We divide the maps into five regions at different galactic latitudes, 
each 10$^o$ wide, and we make pixel-pixel correlations. 
The 410 GHz channel is very well correlated to 
IRAS at low latitudes, and has statistically significant correlations 
for latitudes as high as 50$^o$ (we get linear correlation coefficients
of 0.62, 0.22, 0.20, 0.17, 0.06 for latitude strips centered at 
-15$^o$, -25$^o$, -35$^o$, -45$^o$, -55$^o$ respectively, with about
20000 pixels per strip). The 410 GHz map is morphologically very 
similar to the IRAS/DIRBE map extrapolated at 400 GHz using model 8 
in\cite{Fink}. For $b > -20^o$  the linear correlation coefficient 
between the two datasets is 0.62 (22000 pixels) and the best fit slope 
of the scatter plot is $(1620 \pm 170) MJy/sr / mK_{CMB}$.
The error is dominated by dust properties fluctuations, rather than 
by detector noise. 

The other channels have statistically significant correlations only for low
latitudes ($b > -20^o$). For example, for the 90 GHz channel we have
a correlation coefficient of 0.19 for the strip centered at -15$^o$,
and values $< 0.1$, fluctuating around 0, for higher latitude strips.

We compute the ratio between the dust signal in our channels and the
dust monitored by IRAS at high latitudes ($b < -20^o$)
from the slopes of the best regression line in the pixel-pixel
scatter plots. Only the 410 GHz slope given above is significantly different 
from 0. Using the other slopes we can set upper limits to the contamination 
of dust in our CMB channels, using the mean square fluctuation in the 
IRAS map as a reference. At the 95$\%$ confidence level, the contamination 
at 150 GHz is always less than 3$\%$ of the mean square CMB anisotropy, 
while the contamination of the 90 GHz is always less than 0.5$\%$. 
Needless to say, this analysis does not detect foregrounds which are
not correlated to the 100 $\mu m$ emission mapped by IRAS.

\begin{figure}[!h]
\begin{center}
\includegraphics[height=7.5cm]{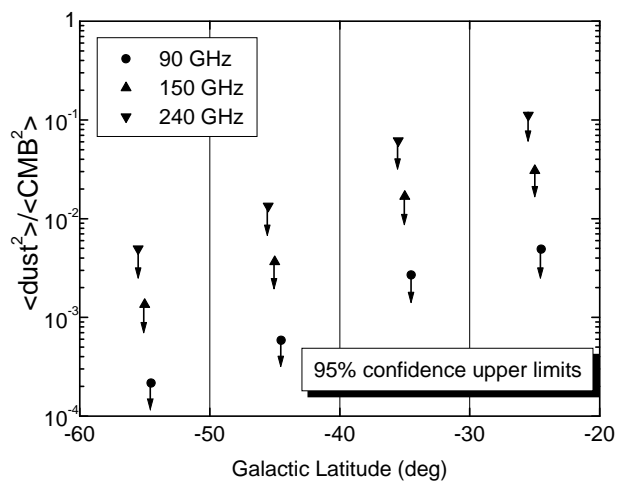}
\caption{Upper limits for the mean square temperature
fluctuations correlated to the IRAS/DIRBE dust template
\label{fig:UL}}
\end{center}
\end{figure}

These are conservative limits, since we are just computing the mean square 
fluctuations, while we know that dust is less and less anisotropic when we 
reduce the angular scale of interest \cite{Gau} \cite{Kog} \cite{Wri}. So the level 
of contamination at $\ell = 200$ can be significantly lower than what we 
have computed here. 

\section{ Conclusions}

These preliminary results confirm that dust is not expected to be a serious
contaminant of sensitive CMB searches, if the region of interest is
a low IRAS brightness one, and at frequencies lower than 170 GHz.
The quality of the data allows us to extract much more.
We are currently working on the determination of dust parameters
and on the estimate of the angular power spectrum of 
fluctuations for this foreground.

\section{ Acknowledgments}

BOOMERanG is supported by PNRA, "La Sapienza", ASI, NSF, NASA, PPARC.
We would like to thank the staff of NSBF (and especially S. E. Peterzen) 
and USAP in Mc Murdo, for the excellent preflight support and the marvelous LDB flight.

\end{document}